# Ghosts of Deletions Past: New Secure Deletion Challenges and Solutions

SARAH DIESBURG, University of Northern Iowa

Is secure deletion of data still a problem?

## 1. INTRODUCTION

The issue of data privacy is becoming increasingly important. While most articles and news pieces concentrate on private information leaked or gathered from the Internet, users must not forget about the difficult issue of preserving privacy on their own electronic devices. In particular, one important aspect of privacy is having control over deletion of data.

The act of completely and irrevocably getting rid of data is called ***secure deletion***. Secure deletion is a difficult action to get right, and unfortunately the advancement of storage devices and operating systems has not made this any easier. With the exception of physical destruction (e.g. smelting, degaussing, acid bath, etc.), software methods may be prone to failure. For example, on-disk secure erase or factory reset commands may report success but actually fail [10] if implemented incorrectly. Some storage devices may not come with any on-disk erase commands at all—think of the thumb drive in the desk drawer! Common device and partition-wide secure deletion programs (such as DBAN [2]) may not erase bad-sector areas on hard drives (HDDs) or non-accessible physical locations on NAND flash devices. Examples of NAND flash devices include thumb drives, solid state drives (SDDs), and smart phone storage.

Full-disk encryption is often touted as a way to confidentially protect data and indirectly solve the secure deletion problem if the device is lost or confiscated. However, this situation is not the same as secure deletion, because the data has not been irrevocably erased. Data can be reinstated if the device is returned to the user. Worse, users of encrypted data may be subjected to coercion for the encryption key, or the device may be subject to decryption or password cracking attempts to restore the password-generated key.

If users would just like to securely delete a few files while keeping the device usable, the field narrows further. Software erasure programs may miss erasure of important metadata, such as file names, file ownership, or access times File system solutions largely rely on the existence of a magnetic-platter hard disk drive and may not erase files from NAND flash devices or parity on RAID systems [6, 7]. Full-disk encryption also cannot work, as the deletion of the single disk key would render all files useless. Using multi-key encryption may work, but only with the added complications of a key store. Of course, encryption keys must also be securely deleted, which again may not be guaranteed due to the problems listed above. This article addresses some of the current challenges surrounding secure deletion and directions for the future.

## 2. SECURE DELETION CHALLENGES

Operating systems exhibit a natural tension with secure deletion. Three main themes emerge in mainstream operating systems:
1. Operating systems are optimized for speed.
2. Operating systems are optimized for reliability.
3. Operating systems are built in layers on the principle of information-hiding (see Fig. 1), so vital information needed for secure deletion is not shared.

Software at the application layer often makes swap and temporary files (reliability mechanism). Secure deletion solutions at this layer can fail because they cannot gain access to file system metadata for deletion (information-hiding mechanism), and they many not delete copies of data made at any lower layer (such as a file system journal, or extra copies of data made on devices which use NAND flash).

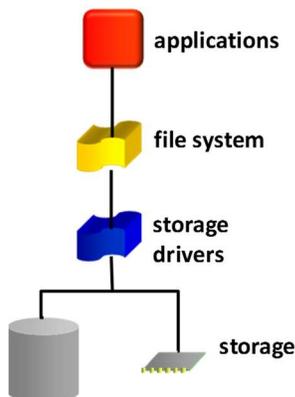

Fig. 1. Operating system storage data path layers.

The file system layer often journals copies of file and metadata information (reliability mechanism). Regular deletion mechanisms at the file system level only cause writes or updates to certain metadata, such as the addition of a deletion flag and an update to a free block bitmap (speed mechanism). General-purpose file systems that have built-in secure deletion commands tend to assume traditional hard drive storage and rely on disk overwrite commands, which may fail for NAND flash storage.

At the driver layer, RAID and other virtual devices often duplicate data (both reliability and speed mechanisms). Further complicating potential secure deletion solutions, the applications and file system layers cannot pass vital information to the driver layer, such as information needed to discern between metadata and file data, or even the notion that a deletion is occurring (information-hiding mechanism). After all, a regular operating system deletion just translates into a series of updates (writes) to metadata, so the lower driver layers only process a series of write commands.

Storage devices themselves also complicate secure deletion. HDDs and SSDs contain bad location lists, and these locations cannot be erased. The user may not be able to see what data are in these lists without resorting to hardware forensics. However, devices which contain NAND flash are even worse. Some NAND flash background may be in order to fully understand the problem.

### 3. NAND FLASH

NAND flash is a type of memory storage with an interface that accepts read, write, and erase requests. NAND has the following characteristics: (1) In-place updates are generally not allowed[1]—once a location is written, the location must be erased before it can be written again; (2) NAND reads and writes are in *flash pages* (e.g., 2–8 KiB), but erasures are performed in *flash blocks* (e.g., 64–512 KiB consisting of contiguous pages); (3) writing is slower than reading, and erasure can be more than an order of magnitude slower; (4) each storage location can be erased only 10K–1M times; and (5) many NAND flash drives have extra non-addressable physical locations that may hold additional information. One study found NAND flash to hold as much as 6%–25% of additional unreported storage area [10].

The challenges posed by these characteristics are commonly solved through the addition of a *flash translation layer (FTL)*. The FTL is usually located in hardware and exports a typical hard drive interface instead of the raw flash interface described above. As a common optimization to mask slow writes and erases, when flash receives a request to overwrite a flash page, the FTL remaps the write to a pre-erased flash page using a translation table, stamps the page (with a version number), and marks the old page as invalid, to be erased later. Information on invalid pages could be read by removing the NAND flash chip and placing it in a universal reader [1]. To prolong the lifespan of the flash, *wear-leveling* techniques are often used to spread the number of erasures evenly across all storage locations. Fig.2 below demonstrates the problem by illustrating typical FTL behavior when a user wishes to overwrite a sensitive area one or more times with patterns and/or random data [4].

---

[1] Flash overwrites might be allowed for some special cases such as marking a page invalid.

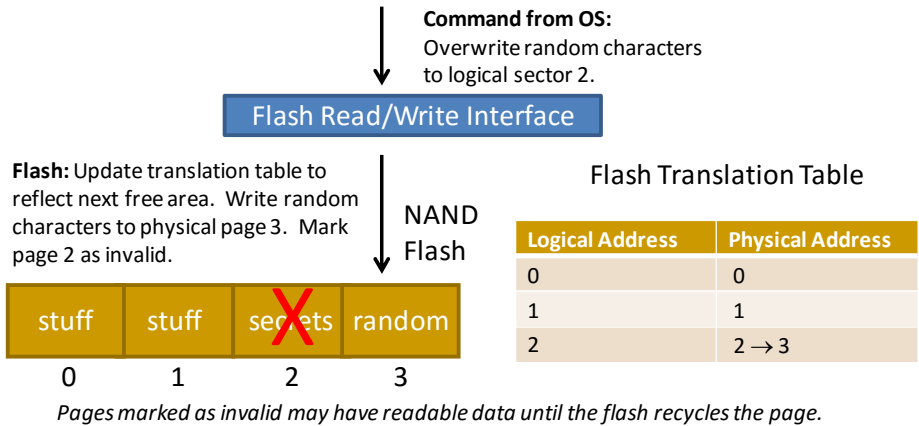

*Pages marked as invalid may have readable data until the flash recycles the page.*

Fig 2. Illustration of an attempted secure deletion using overwrites.

If a partition- or disk-wide overwriting program like DBAN is used, the problem still persists due to the extra storage areas built into the NAND flash. Overwritten pages are marked invalid and may be moved into the inaccessible, extra page range to speed wear-leveling before erasure. Even parts of files overwritten during normal file modification may wind up in the extra page range. Due to the proprietary nature of the FTL algorithms, it is not known exactly when those pages will be erased.

## 4. FUTURE DIRECTIONS

The secure deletion problem cannot feasibly be solved through increased user awareness. For example, users (even non-technical ones) must be aware of the physical characteristics of their storage devices, complications imposed by their file systems, and limitations imposed by software solutions to choose the best option from the myriad array of commercial, free, and built-in secure deletion products available.

Perhaps the secure deletion problem is instead an engineering problem. The current user model is to make users expend effort and use often advanced tools to securely delete files. *If the default deletion behavior is secure deletion, then users would have to use extra effort to turn secure deletion off.*

Storage encryption may be able to solve the secure deletion problem in the circumstance that the encryption key is completely destroyed, the encryption method is strong, and the implementation of the encryption does not contain errors or back doors. If keys are saved on the storage, then they also must be securely erased.

Operating systems could be redesigned to make secure deletion easier. One step in the right direction is to enable the file system to pass an erasure command all the way down to the storage driver [3, 8]. The NAND flash TRIM command [9] illustrates this full-data-path behavior by allowing the file system to specify blocks no longer in use so that the FTL may garbage collect them at a later time. In its current form, TRIM is implemented for SSD performance.

If a command like TRIM could be adapted for secure deletion, its implementation and use would need some modifications. (1) Deletion must happen immediately, or else, at a guaranteed, verifiable time in the near future. Also, all pages that hold copies of this data must be tracked down and securely deleted. An example of a stack-based FTL that automatically tracks old versions of pages/blocks is INFTL, and such an FTL design could be leveraged for secure deletion [3]. To decrease device wear, an encryption key could be securely deleted in lieu of data [8], or only certain files could be securely deleted (such as in a user's home directory). (2) Metadata must also be deleted, including metadata in a file system journal. Currently, TRIM-enabled file systems tend to concentrate on data blocks. (3) Other storage devices (like HDDs) should also recognize the command and implement suitable deletion.

The new eMMC storage sanitize command [5] is a step in the right direction and will remove data from the unmapped eMMC data regions. However, unless verification of secure deletion is possible, this solution and the above proposed solutions are moot. *Security by obscurity should not be the norm in our increasingly-complex storage devices.* One solution may be to create open and verifiable storage so that all contents on storage could be viewed and encryption algorithms (if any) could be verified. The Linux Memory Technology Device (MTD) subsystem contains software FTL support and access to the raw storage areas through a character device. Drives with hardware FTLs, such as SSDs, could be extended to export a similar, read-only interface to view the raw contents of data areas on the SSD without exposing proprietary microcode or FTL management areas.

In summary, secure deletion is still a hard problem primarily due to operating system and storage design. However, it is a solvable problem for designers of operating systems and storage devices if (1) deletion behavior defaults to secure deletion, (2) operating systems are designed to share information about deletions to storage drives, and (3) storage devices are designed in an open and verifiable manner.